\documentclass[twocolumn,showpacs,aps,draft]{revtex4}

\usepackage[dvips,final]{graphics}
\usepackage[dvips]{graphicx}
\usepackage{bm}
\usepackage{amsmath}

\begin{document}

\newcommand{\no}{\nonumber}
\newcommand{\etal}{{\em et~al }}
\newcommand{\ie}{{\it i.e.\/}\ }
\newcommand{\smallsection}[1]{{\par\it #1.---}}
\newcommand{\void}{1}{}

\title{\center{A multimode model for projective photon-counting measurements}}

\author{Rosa Tualle-Brouri$\,$}
\email[E-mail:$\;\;$]{rosa.tualle-brouri@institutoptique.fr}
\author{Alexei Ourjoumtsev}
\author{Aurelien Dantan}
\author{Philippe Grangier}
\affiliation{Laboratoire Charles Fabry de l'Institut d'Optique,
CNRS UMR 8501, Universit\'e Paris Sud XI, 91127 Palaiseau, France
}
\author{Martijn Wubs${\,}^{\dag,\mathsection}$}
\author{Anders S. S{\o}rensen${\,}^{\mathsection}$}
\affiliation{${}^\dag$  Niels Bohr International Academy $\& $ ${}^\mathsection$ QUANTOP  \\
 The Niels Bohr Institute, Blegdamsvej 17, DK-2100 Copenhagen, Denmark}
\date{Version 17, date: \today}

\begin{abstract}

We present a general model to account for the multimode nature of
the quantum electromagnetic field in projective photon-counting
measurements. We focus on photon-subtraction experiments, where
non-gaussian states are produced conditionally. These are useful
states for continuous-variable quantum information processing.  We
present a general method called mode reduction that reduces the
multimode model to an effective two-mode problem. We apply this
method to a multimode model describing broadband parametric
downconversion, thereby improving the analysis of existing
experimental results. The main improvement is that spatial and
frequency filters before the photon detector are taken into
account explicitly. We find excellent agreement with previously
published experimental results, using fewer free parameters than
before, and discuss the implications of our analysis for the
optimized production of states with negative Wigner functions.

\end{abstract}

\maketitle

\section{Introduction}

The ability to prepare and measure specific quantum states of the
light is the keystone of many quantum information processing (QIP)
protocols. These states can be described either with discrete
variables in terms of photons, or with continuous variables in
terms of waves. In the latter case, the physical quantities of
interest are the amplitude and the phase of the light wave, or
their Cartesian counterparts called quadratures $\widehat{x}$ and
$\widehat{p}$. A very convenient representation of the quantum
state is then provided by the Wigner function $W(x,p)$, which
corresponds to a quasi-probability distribution of the
quadratures, `quasi-' because $W$ may assume negative values.

An important task for QIP is the ability to undo effects of
decoherence by `distillation': to obtain a single quantum state
that is more pure from two or more copies that have undergone
decoherence. Since states of light with gaussian Wigner functions
cannot be distilled with gaussian operations
\cite{Eisert2002a,Giedke}, one is left with two  strategies:
either to distill gaussian states with non-gaussian operations, or
to distill non-gaussian states with gaussian operations
\cite{Browne}.  This paper is a contribution to the latter
strategy, and focuses on the preparation of the non-gaussian
states rather than on their distillation.

The negativity of the Wigner function is a standard figure of
merit, quantifying at the same time how non-gaussian and how
non-classical a quantum state is~\cite{Kim2005a,Biswas2007a}.
One way of obtaining non-gausian states is by conditional
photon-counting measurements, as first proposed by Dakna {\em et
al.}~\cite{Dakna1997a}. It was soon realized that such conditional
measurements can improve quantum teleportation of continuous
variables~\cite{Opatrny2000a}.
In recent years, several
experiments~\cite{Lvovsky,WengerCond,Ourjoumtsev2006a,Ourjoumtsev2006b,Wakui,Neergaard,Ourjoumtsev2006c,Parigi2007a}
combining continuous- and discrete-variable tools allowed for
preparing and observing quantum states of free-propagating light
with negative Wigner functions~\cite{Kim2008a}.

Many of these experiments are based on the use of a squeezed
vacuum produced by parametric fluorescence, which involves many
optical modes~\cite{Avenhaus}. This multimode nature is exhibited
in both continuous-wave (CW) operation,  using optical parametric
oscillators (OPO) below threshold~\cite{Suzuki,Molmer}, and in
pulsed experiments with a single-pass high amplification.
In order to make accurate predictions, it is therefore crucial to
develop multimode theoretical models. This was done
in~\cite{Suzuki,Molmer,Nielsen2007a, Nielsen2007b} for setups
using an OPO, and in~\cite{prl,Aichele} for pulsed $1$-photon Fock
state tomography in a case of very low squeezing, when Fock states
expansion are limited to $1$ photon only. However, these models do
not fully account for all phenomena linked to the non-constant
space and time profiles of the modes under study, the spatial
pulse profile in the transverse direction for example, and they
especially do not account for gain-induced distortions in the
parametric amplification process~\cite{LaPorta}. As we will see
later on, these phenomena are one main signature of this multimode
nature, and are critical in the case of single-pass pulsed
experiments.

In this paper we propose an alternative general framework to
describe the generation of squeezed light, with a twofold goal:
first, to introduce a method that reduces a multimode model to an
effective two-mode description. Second, to show that a specific
spatio-temporal multimode model for photon-subtraction experiments
and our mode-reduction analysis thereof give an improved
understanding of state-of-the-art photon-subtraction experiments.

In Sec.~\ref{secModered} we show how to reduce a multi-mode model
to an effective two-mode model. This mode-reduction procedure is
then applied in Sec.~\ref{secApp} to give an improved analysis of
the photon-subtraction experiments of
Ref.~\cite{Ourjoumtsev2006a}. We discuss the method and its
application and conclude in Sec.~\ref{secDiscConc}. Some
technicalities are deferred to two Appendices.

\section{Reduction of multimode model}\label{secModered}

\subsection{General multimode model}\label{secMultimodemodel}
 As a starting point we have a complete set of spatial, temporal, or spatio-temporal optical modes, in terms of which
the light propagation can be described. The modes have field
operators $\hat{\text{a}}_j$ and $\hat{\text{a}}_j^{\dag}$ with
bosonic commutation relations
$[\hat{\text{a}}_j,\hat{\text{a}}_k^\dag]=\delta_{jk}$. The
$\hat{\text{a}}_{j}$ may also stand for continuous operators like
$\hat{\text{a}}(t)$, where $t$ is time, with commutation relation
$[\hat{\text{a}}(t),\hat{\text{a}}^{\dag}(t')]=\delta(t-t')$, in
which case sums over modes are to be replaced with integrals. To
simplify the notation even further, we introduce
$\vec{\hat{\text{a}}}$, the column vector of all the
$\hat{\text{a}}_j$. Similarly $\vec{\hat{\text{a}}}^\dag$ is the
row vector which has operators $\hat{\text{a}}_j^\dag$ as
components. We shall also need the row vector
$\vec{\hat{\text{a}}}^T$ and the column vector
$\vec{\hat{\text{a}}}^*=\vec{\hat{\text{a}}}^{\dag T}$. Similar
notation will be used for other vectors.

\begin{figure}[t]
\center
\includegraphics{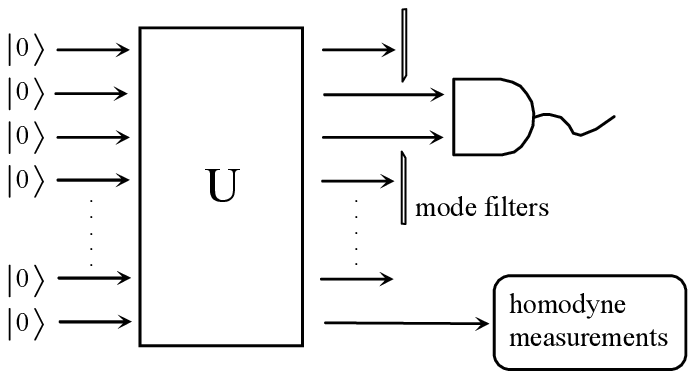}
 \caption{Projective photon-counting
measurement: in our model we assume that we start with a multimode
vacuum input state which is subject to a general multimode
Bogoliubov transform $U$. After the transformation a few modes are
filtered out and result in a detection event in the avalanche
photodiode (APD). Such a detection event in the APD prepares a
state in the single mode which is mode analyzed by homodyne
detection.} \label{genSetup}
\end{figure}

In this section we consider a general unitary transformation $U$
in which output quadratures linearly depend on input quadratures,
thus preserving the gaussian nature of the quantum fields. In
fact, the only non-gaussian operation will be the projective
measurement, corresponding to a detection event in a subset of the
output modes by an avalanche photodiode (APD), see
Figure~\ref{genSetup}.
We will assume the evolution operator $U$ in Fig.~\ref{genSetup}
to be
 a Bogoliubov transformation, where the output field
operators  depend linearly on the input ones:
\begin{equation} \label{Bogoliubov}
\qquad\vec{\hat{\text{a}}}_{\rm out}=U^\dag
\vec{\hat{\text{a}}}\,U=
u\,\vec{\hat{\text{a}}}+v\,\vec{\hat{\text{a}}}^*,
\end{equation}
where $u$ and $v$ are two matrices which satisfy
\begin{subequations}
\begin{eqnarray}\label{coef}
uu^\dag-vv^\dag & = & u^\dag u-v^\dag v = 1,\\
uv^T-vu^T & = & 0,
\end{eqnarray}
\end{subequations}
whereby the commutation relations of the field operators are
preserved. The output state of the light can be characterized by
doing homodyne measurements on a normalized mode described by
$\vec{\psi}_h$, which has the mode operator
\begin{eqnarray}\label{ah}
\hat{a}_h=\vec{\psi}_h^\dag \,\vec{\hat{\text{a}}}.
\end{eqnarray}
This mode can be defined as the mode that perfectly matches the
local oscillator of the homodyne detector. After the Bogoliubov
transform~(\ref{Bogoliubov}) it is described by
\begin{eqnarray}\label{ahout}
\hat{a}_{h,{\rm out}}=U^\dag \hat{a}_h U=\vec{\psi}_h^\dag
(u\,\vec{\hat{\text{a}}}+v\,\vec{\hat{\text{a}}}^*).
\end{eqnarray}
Without conditioning upon photon detection, and since the gaussian
nature of the initial vacuum state is preserved by the Bogoliubov
transform, the homodyne measurements will show gaussian Wigner
functions corresponding to squeezed vacuum states of light.

The point is now how to describe the output state conditional upon
detection of a photon by the APD, which is a projective
measurement. Analogous to the homodyne detection, we assume the
photon detection mode $j$ to be described by a normalized state
$\phi_d(j)$, with corresponding field operator
\begin{eqnarray}\label{ad}
\hat{a}_d(j)=\vec{\phi}_d^\dag(j) \vec{\hat{\text{a}}}.
\end{eqnarray}
After the time evolution described by Eq.~(\ref{Bogoliubov}), the
output at the photon detector becomes:
\begin{eqnarray}\label{adout}
\hat{a}_{d,{\rm out}}(j)=U^\dag \hat{a}_d(j)
U=\vec{\phi}_d^\dag(j)
(u\,\vec{\hat{\text{a}}}+v\,\vec{\hat{\text{a}}}^*).
\end{eqnarray}
If we only know that a photon has been detected but not in which
detection mode, then we should average the conditional output
state over all these detection modes, as is shown in more detail
below.

At this point we have a multimode output state that in principle
we can update given the detection of a photon in the detector $j$.
In practice it is convenient to first simplify the multimode
expressions~(\ref{ahout}) and (\ref{adout}).


\subsection{Mode reduction}\label{SecModered}
The first and central step in the mode-reduction procedure is to
rewrite the homodyne mode Eq.~(\ref{ahout}) in the form
\begin{equation}
\hat{a}_{h,{\rm out}}=\eta\, \hat{a}_0+\alpha\,
\hat{a}_0^\dag+\beta\,\hat{a}_1^\dag\label{modereductionahout},
\end{equation}
in terms of new mode operators $a_{0,1}^{(\dag)}$ with standard
commutation relations. The coefficients $\eta,\alpha$, and $\beta$
in Eq.~(\ref{modereductionahout}) are found as follows. Besides
having standard commutations, $\hat{a}_0$ and $\hat{a}_1$ in
Eq.~(\ref{modereductionahout}) should annihilate the vacuum state,
so that $\hat{a}_0$ must contain all annihilation operators in
Eq.~(\ref{ahout}):
\begin{figure}[t]
\center
\includegraphics{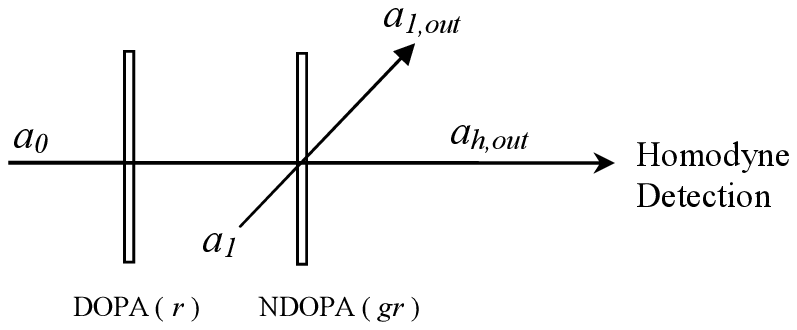}
 \caption{Equivalent model for the Bogoliubov
transform U: a perfect single-mode degenerate optical parametric
amplifier (DOPA) with squeezing parameter $r$ is followed by a
perfect non-degenerate optical parametric amplifier (NDOPA) with
squeezing parameter $gr$. } \label{modelU}
\end{figure}
\begin{eqnarray}\label{a0}
\eta\,\hat{a}_0=\vec{\psi}_h^\dag u\,\vec{\hat{\text{a}}},
\end{eqnarray}
with $\eta$ fixed up to a phase factor by
$[\hat{a}_0,\hat{a}_0^\dag]=1$. We choose $\eta$ to be real-valued
and positive so that
\begin{eqnarray}\label{eta}
\eta=\sqrt{\vec{\psi}_h^\dag uu^\dag \vec{\psi}_h}.
\end{eqnarray}
Furthermore, from $[\hat{a}_0,\hat{a}_{1}^{\dag}]=0$ it follows
that
\begin{eqnarray}\label{alpha}
\alpha=[\hat{a}_0,\hat{a}_{h,{\rm
out}}]=\frac{1}{\eta}\vec{\psi}_h^\dag \,u\,v^T\, \vec{\psi}_h^*.
\end{eqnarray}
A complex value for $\alpha$ can be removed with a redefinition of
the phase of the homodyne mode, which just means that we can
assume that the state is squeezed in the $x$ or $p$ direction.
Finally, $\beta$ can be found from the fact that $\hat{a}_{h,{\rm
out}}$ and $\hat{a}_{h,{\rm out}}^\dag$ have commutator one, as
have $\hat{a}_{1}$ and $\hat{a}_{1}^\dag$:
\begin{eqnarray}\label{beta}
\beta=\sqrt{\eta^2-\alpha^2-1}.
\end{eqnarray}
Here again we used the freedom to choose $\beta$ real-valued and
nonnegative. This completes the mathematics of the mode reduction
of the multimode homodyning signal.

The physical argument that two and only two modes should remain
goes as follows. The squeezed vacuum after the Bogoliubov
transform can only be a centered gaussian state, hence it is fully
described by only the variances $V_x$ and $V_p$. The squeezed
vacuum output can therefore be modeled~\cite{Gardiner2000a,Wenger}
by a perfect single-mode degenerate optical parametric amplifier
(DOPA) with squeezing parameter $r$, followed by a perfect
non-degenerate optical parametric amplifier (NDOPA) with squeezing
parameter $g r$, as presented in Figure~\ref{modelU}. This is a
two-mode model, with a Hilbert space $\mathcal{H}_{2}$. The values
of the parameters $r$ and
 $g $ can be deduced from the two independent coefficients in
 Eqs.~(\ref{eta}-\ref{beta}), using:
\begin{subequations}\label{equ_gr}
\begin{eqnarray}
\eta & = & \cosh(r)\cosh(gr) \\
\alpha & = & \sinh(r)\cosh(gr) \\
\beta & = & \sinh(gr).
\end{eqnarray}
\end{subequations}
A related squeezing parameter that we will also use in the
following is $s \equiv \exp(- 2 r)$.

\subsection{Conditioning upon photon detection}\label{SecConditioning}
We now condition upon the measurement of a click in the photon
detector (APD). We assume to be in the limit that the average
number of photons per pulse entering the photon detector is much
less than one. Then a single click in the detector corresponds to
the detection of a single photon.

One can make a reduced-mode description of the photon detection
operator $\hat{a}_{d,{\rm out}}(j)$ of Eq.~(\ref{ad}), analogous
to Eq.~(\ref{modereductionahout}). The operator $\hat{a}_{d,{\rm
out}}(j)$ can be expanded into a part acting on $\mathcal{H}_{2}$,
plus a component $\hat{a}_{d\perp}(j)$ acting on the complementary
space $\mathcal{H}_\perp$ orthogonal to $\mathcal{H}_{2}$:
\begin{eqnarray}\label{adperp_Mode_Red}
\hat{a}_{d,{\rm
out}}(j)=\gamma_j\hat{a}_0^\dag+\delta_j\hat{a}_1^\dag
+\epsilon_j\hat{a}_0+\kappa_j\hat{a}_1+\hat{a}_{d\perp}(j).
\end{eqnarray}
Note that $\hat{a}_{d\perp}(j)$ on the right-hand side contains
all the terms acting on $\mathcal{H}_\perp$, including creation
operators. The coefficients in~(\ref{adperp_Mode_Red}) can again
be found by taking commutators, for example:
\begin{subequations}\label{gammadelta}
\begin{eqnarray}
\gamma_j & = & [\hat{a}_0,\hat{a}_{d,{\rm
out}}(j)]=\frac{1}{\eta}\vec{\psi}_h^\dag
\,u\,v^T\, \vec{\phi}_d^*(j) \label{gamma}\\
\delta_j & = & [\hat{a}_1,\hat{a}_{d,{\rm
out}}(j)]=\frac{1}{\beta^*}[\vec{\phi}_d^\dag(j) \,v\,v^\dag\,
\vec{\psi}_h-\alpha^*\gamma_j] \label{delta}.
\end{eqnarray}
\end{subequations}

In general, a click recorded in  the APD corresponds to the
measurement of at least one photon. In the limit of low detection
probability, the action of the detection is the subtraction of a
{\em single} photon. Note that this assumption is practically
always obeyed if the parameter $j$ is in a continuum (like in the
case of spectral filtering), since the probability to have two
photons exactly in the same mode is then negligible. Henceforth we
assume to be in this limit. In the Heisenberg picture, a photon
detection then corresponds to the application of the operator
$\hat{a}_{d,{\rm out}}(j)$ to the initial state (\ie to the vacuum
state), followed by the normalization of the result:
\begin{eqnarray}\label{PsiCond}
|\Psi(j)\rangle=P_j^{-1/2}\hat{a}_{d,{\rm out}}(j)|0\rangle
\end{eqnarray}
where $P_j$ is the detection probability for the mode $j$:
%
\begin{eqnarray}
P_j & =& \langle0|\hat{a}_{d,{\rm out}}^\dag(j)\hat{a}_{d,{\rm out}}(j)|0\rangle \nonumber \\
& =& \langle0|\hat{a}_{d\perp}^\dag(j)\hat{a}_{d\perp}(j)|0\rangle+|\gamma_j|^2+|\delta_j|^2 \label{Pj}\\
& =& \overrightarrow{\phi}_d^\dag(j) v v^\dagger
\overrightarrow{\phi}_d(j). \nonumber
\end{eqnarray}
Below we will use that the vacuum expectation value of $\hat
n_{d\perp}(j) = \hat{a}_{d\perp}^\dag(j)\hat{a}_{d\perp}(j)$ can
be expressed as $P_j - |\gamma_j|^2-|\delta_j|^2$.

Before continuing, it can be instructive to recall the concision
allowed by the Heisenberg picture. In a single-mode problem, a
photon-subtracted squeezed state is equivalent to a squeezed
single-photon state: this case corresponds to the simple
Bogoliubov transform $\hat{a}_{\rm out}=\cosh(r) \hat{a}_{\rm
in}+\sinh(r) \hat{a}^\dag_{\rm in}$, which directly gives a pure
$1$-photon state (after normalization) when applied to the vacuum.
As the states do not evolve in the Heisenberg picture, they all
can be considered as `input' states; but when measured using
output quadratures, this $1$-photon state will appear to be
squeezed.

We are going to use the same approach in the multimode case. One
can first note that the conditioned state $|\Psi(j)\rangle$  in
Eq.~(\ref{PsiCond}) is already a $1$-photon state. This state,
however, does not belong to $\mathcal{H}_{2}$ only, so that
measurements output are not so obvious to compute. In fact, we are
solely interested in expectation values $\langle g(a_{h,\rm
out},a_{h,\rm out}^{\dag})\rangle$ of operators describing the
output that is measured in the homodyne detector. Such expectation
values can be written as
\begin{subequations}
\begin{eqnarray}\label{rewriteexpval}
\langle g(a_{h,\rm out},a_{h,\rm out}^{\dag})\rangle & = & \langle
\psi_j|\, g(\hat a_{h, {\rm out}},\hat a_{h, {\rm
out}}^{\dag})\,|\psi_j\rangle \\
& = & \mbox{Tr}\{g(\hat a_{h, {\rm out}},\hat a_{h, {\rm
out}}^{\dag})|\psi_j\rangle\langle \psi_j|\}.~~~~~~
\label{rewriteexpvalb}
\end{eqnarray}
\end{subequations}
In Eq.~(\ref{rewriteexpvalb}), the trace can be separated into a
trace over $\mathcal{H}_{2}$ and a trace over
$\mathcal{H}_{\perp}$, and the latter does not act on the function
$g(\hat a_{h, {\rm out}},\hat a_{h, {\rm out}}^{\dag})$, whose
expectation value is then:
\begin{subequations}
\begin{eqnarray}
 & = &  \mbox{Tr}_2\{g(\hat
a_{h, {\rm out}},\hat a_{h, {\rm
out}}^{\dag})\mbox{Tr}_\perp|\psi_j\rangle\langle \psi_j|\} \\
& = &  \mbox{Tr}_2\{g(\hat a_{h, {\rm out}},\hat a_{h, {\rm
out}}^{\dag})\rho_j\}\label{rewriteexpvald}
\end{eqnarray}
\end{subequations}
All quantities of interest can therefore be deduced from the input
state reduced density matrix $\rho_j$, acting in
$\mathcal{H}_{2}$, and the crucial advantage of mode reduction is
to allow a simple expression for this matrix: Writing
$|0\rangle=|00\rangle\otimes|0\rangle_\perp$, where $|00\rangle$
and $|0\rangle_\perp$ are the ground states of $\mathcal{H}_{2}$
and $\mathcal{H}_\perp$, respectively, and using
Eqs.~(\ref{adperp_Mode_Red},\ref{PsiCond}), we directly obtain:
\begin{eqnarray}
\rho_j & = & \mbox{Tr}_{\perp}|\psi_{j}\rangle\langle\psi_{j}|  \label{rho2} \\
& = & (1-\xi_j)\,|00\rangle\langle 00|+\xi_j \,
\hat{a}_\theta^\dag(j)|00\rangle\langle 00|\hat{a}_\theta(j),
\nonumber
\end{eqnarray}
in terms of the {\em modal purity}
\begin{eqnarray}\label{xij}
\xi_j=\frac{|\gamma_j|^2+|\delta_j|^2}{P_j},
\end{eqnarray}
and where
\begin{equation}\label{athetaj}
\hat{a}^\dag_{\theta}(j)=\cos\theta_j \hat{a}_0^\dag+\sin\theta_j
\hat{a}_1^\dag, \quad \mbox{with}\;\;
\tan\theta_j=\delta_j/\gamma_j
\end{equation}
is an operator that creates a single photon in a superposition of
mode~0 and mode~1. The state~(\ref{rho2}) produced from a
detection event in mode $j$ is a mixed state, mixing vacuum and a
single-photon state with weight $\xi_j$. Without conditioning or
in the limit $\xi_{j} \rightarrow 0$, we have
$\rho_{j}=|00\rangle\langle 00|$.

\subsection{Wigner functions}\label{secWigner}
\smallsection{Squeezed vacuum} Before determining the output
Wigner function corresponding to the conditional
state~(\ref{rho2}), it is instructive to first determine the
Wigner function of the output state in the simplest experimental
situation, where we ignore the photon detector. The input state is
then $\rho = |00\rangle\langle 00|$. We will make use of the
standard Wigner functions of the vacuum $W_{0}(x,p) =
\exp(-r^{2})/\pi$ and of single-photon states $W_{1}(x,p) =
(2r^{2}-1)\exp(-r^{2})/\pi$, both with $r^{2} = x^{2}+p^{2}$.
Clearly, $W[\rho](x_{0},p_{0};x_{1},p_{1})$ equals
$W_{0}(x_{0},p_{0})W_{0}(x_{1},p_{1})$.

In order to obtain the output Wigner function for the homodyne
mode, we wish to express $W[\rho]$ as a function of $x_{h,\rm
out}$, $p_{h,\rm out}$ (defined from the output homodyne mode
$\hat{a}_{h,\rm out}$ given by Eq.~(\ref{modereductionahout})).
This requires the introduction of another mode $\hat{a}_{1,{\rm
out}}$ orthogonal to $a_{h,{\rm out}}$, so that the transformation
$(\hat{a}_0,\hat{a}_1)\rightarrow(\hat{a}_{h,\rm
out},\hat{a}_{1,\rm out})$ is symplectic (\ie commutation
relations are preserved). Using the model of Fig.~\ref{modelU},
one can choose $\hat{a}_{1,{\rm out}}$ of the form:
\begin{eqnarray}\label{a1out}
\hat{a}_{1,{\rm out}}=\beta\,\frac{\alpha\, \hat{a}_0+\eta\,
\hat{a}_0^\dag}{\sqrt{1+\beta^2}}+\sqrt{1+\beta^2}\, \hat{a}_1.
\label{DOPAperp}
\end{eqnarray}
This form is by no means unique, but this does not pose a problem
since mode $1_{\rm out}$ will eventually be integrated out. One
can now invert the relations
(\ref{modereductionahout},\ref{a1out}), thereby expressing
$x_{0,1},p_{0,1}$ as a function of $x_{h,{\rm out}}, p_{h,{\rm
out}}, x_{1,{\rm out}}$, and $p_{1,{\rm out}}$. After tracing over
mode $1_{\rm out}$, which amounts to integrating over $x_{1,{\rm
out}}$ and $p_{1,{\rm out}}$, we find the output signal entering
the homodyne detector to be a squeezed vacuum state with a
gaussian Wigner function
\begin{equation} \label{0_sqz}
W_{0,{\rm sqz}}(x,p)=\frac{1}{\pi\sqrt{V_x
V_p}}\exp\left(-\frac{x^2}{V_x}-\frac{p^2}{V_p}\right),
\end{equation}
where $x$ and $p$ stand for $x_{h,{\rm out}}$ and $p_{h,{\rm
out}}$ and with variances
\begin{subequations}\label{defVar}
\begin{eqnarray}
V_{x} &= & (\eta + \alpha)^2+\beta^2 \\
V_{p} & =& (\eta - \alpha)^2+\beta^2.
\end{eqnarray}
\end{subequations}
We can now invert Eq.~(\ref{defVar}) and rewrite the three
mode-reduction parameters $\alpha,\eta$, and $\beta$ in
Eq.~(\ref{modereductionahout}) in terms of the variances, giving:
\begin{subequations}\label{etaalphabetaVxVp}
\begin{eqnarray}
\eta & = &\frac{V_{p}+V_{x}+2}{2 \sqrt{V_{x}+V_{p}+2}},   \\
\alpha & = & \frac{V_{x}-V_{p}}{2 \sqrt{V_{x}+V_{p}+2}},  \\
\beta & = & \frac{2 \sqrt{V_{x}V_{p}-1}}{2 \sqrt{V_{x}+V_{p}+2}}.
\end{eqnarray}
\end{subequations}
In the following, we will keep writing $\alpha,\eta$, and $\beta$
to shorten notation. It should be kept in mind, however, that
Eq.~(\ref{etaalphabetaVxVp}) directly expresses these parameters
in terms of the measurable variances $V_{x,p}$ of the squeezed
vacuum. In particular, $\beta$ vanishes for minimal-uncertainty
states.

Notice also that the parametrization for the mode-reduction
parameters~(\ref{etaalphabetaVxVp}) is equivalent to the one in
Eq.~(\ref{equ_gr}) in terms of squeezing parameters $r$ and $g$.
Thus $r$ and $g$ can be expressed in terms of the variances
$V_{x,p}$, and {\em vice versa}.

\smallsection{Photon-subtracted squeezed vacuum} As for the
squeezed vacuum, we now calculate the Wigner function for the
photon-subtracted squeezed vacuum, starting with the initial
state~(\ref{rho2}).
The mode~(\ref{athetaj}) has a one-photon excitation in
state~(\ref{rho2}). The orthogonal mode with creation operator
$\hat a_{\theta+\pi/2}^{\dag}(j)$ is not excited. Hence the Wigner
function corresponding to the state~(\ref{rho2}) is
\begin{eqnarray}\label{Winitial}
W(x_{0},p_{0};x_{1},p_{1})&  = & (1-\xi_j)\,
W_{0}(x_{0},p_{0})W_{0}(x_{1},p_{1})  \\ & + & \xi_{j}\,
W_{1}(x_{\theta_j},p_{\theta_j})W_{0}(x_{\theta_j+\pi/2},p_{\theta_j+\pi/2}).
\nonumber
\end{eqnarray}
Note that in this expression, quadratures
$x_{\theta_j,\theta_j+\pi/2}$, $p_{\theta_j,\theta_j+\pi/2}$ can
be easily expressed as functions of quadratures $x_{0,1},p_{0,1}$
using (\ref{athetaj}). As before, the Wigner function for the
output signal is found by using the symplectic transformation
defined by Eqs.~(\ref{modereductionahout}) and (\ref{a1out}). By
tracing again over the mode $1_{\rm out}$,  we obtain (see
appendix \ref{AppendixWigner})
\begin{equation} \label{Wj}
W_j(x,p)=\left(C_j+2A_j\frac{x^2}{V^2_x}+2B_j\frac{p^2}{V^2_p}+D_j\frac{xp}{V_xV_p}\right)
W_{0,{\rm sqz}}.
\end{equation}
The constants in this Wigner function are given by
\begin{subequations}\label{constantsinW}
\begin{eqnarray}
A_j & = & P_j^{-1}|\gamma_j(\eta+\alpha)+\delta_j\beta|^2\label{Aj},\\
B_j & = & P_j^{-1}|\gamma_j(\eta-\alpha)-\delta_j\beta|^2\label{Bj},\\
C_j & = & 1-A_j/V_x-B_j/V_p, \label{Cj} \\
D_j & = & -8
P_j^{-1}\mbox{Im}(\gamma^*_j\delta_j)\eta\beta\label{Dj}.
\end{eqnarray}
\end{subequations}
 Note that the Wigner function~(\ref{Wj}) of the
photon-subtracted squeezed state differs from the Wigner function
of the squeezed vacuum  $W_{0,{\rm sqz}}(x,p)$ of
Eq.~(\ref{0_sqz}) only because of the polynomial in $x$ and $p$
between the large brackets. The same quantities $V_{x,p}$ as in
Eq.~(\ref{defVar}) show up, with or without conditioning. Since in
general $W(x,p)\ge - \pi^{-1}$ \cite{Leonhardt1997a}, we find the
condition $C_{j}\ge -1$.

\smallsection{Averaged Wigner functions} Practical detectors do
not resolve with infinite precision when and where photons are
detected. We should therefore average over all possible
microscopic states that agree with the detection record.
  We assumed in Sec.~\ref{SecConditioning} that the average number of photons detected per pulse in the APD is much smaller than one.
  Averaging over unresolved detection events is then equivalent to averaging over single-photon subtraction events.

The Wigner transformation of the density matrix is a linear
transformation. Therefore, the averaged Wigner function
$\overline{W}(x,p)$ is simply obtained by replacing $A_j$...$D_j$
in~(\ref{Wj}) by $\overline{A}$...$\overline{D}$, with the
notation
\begin{equation} \label{Xbar}
\overline{X}=P_{\rm tot}^{-1}\sum_{j} P_{j}X_{j}.
\end{equation}
Here $P_{\rm tot}$ is the sum of the probabilities $P_j$ of
microscopic states that agree with the detection record. From
Eq.~(\ref{constantsinW}) it follows that averaged quantities
$\overline{A}$...$\overline{D}$ involve sums like $\sum_j
|\gamma_j|^2$, $\sum_j |\delta_j|^2$ or $\sum_j
\gamma_j^*\delta_j$; in the following we will write the respective
averages as $\overline{|\gamma|^2}$, $\overline{|\delta|^2}$ or
$\overline{\gamma^*\delta}$.

The Wigner function~(\ref{Wj}) of the photon-subtracted state and
its detection-averaged version have a very general significance.
Before going to Sec.~\ref{secApp}, devoted to a practical
implementation of these results, let us finish with some
reflections on the detection modes.

\subsection{Detection modes}\label{subsecDetModes}

All the previous results were derived through the use of a set of
detection modes $\hat{a}_d(j)$. The coefficients entering in the
averaged Wigner function $\overline{W}$ involve quantities like
$P_{\rm tot}$, $\overline{|\gamma|^2}$, $\overline{|\delta|^2}$ or
$\overline{\gamma^*\delta}$, in which the detection operators only
appear through their projection operator:
\begin{equation} \label{equProjP}
\Pi=\sum_j\vec\phi_d(j)\vec\phi_d^\dag(j).
\end{equation}
Hence one would find the same predicted averages if one would
employ a different set of detection modes that has the same
associated projector. This projector describes how the setup
filters the signal before it enters the photon detector. For
example, a time-domain filtering system will be described by
$\Pi=\Pi_T$, where $\Pi_T$ can be written in terms of a set of
modes $\hat{\text{a}}(t)$ that are labeled by the time $t$:
\begin{equation} \label{equProjPT}
\Pi_T(t,t')=T(t)\delta(t-t'),
\end{equation}
where $T(t)=1$ when the APD is switched on, and $T(t)=0$
otherwise. To give another important example, a spectral slit can
be described with a set of modes $\hat{\text{a}}(\omega)$ with an
associated projector
\begin{equation} \label{equProjPOmega}
\Pi_\Omega(\omega,\omega')=T(\omega)\delta(\omega-\omega'),
\end{equation}
where $T(\omega)=0$
 if the frequency $\omega$ is filtered out, and $T(\omega)=1$ otherwise.

Above we have assumed that the filtering of the signal after its
production in the DOPA was included in the Bogoliubov transform
$U$. Below we will give an alternative description, in which $U$
is separated into the transformation due to the production of the
squeezed light in the DOPA, and the subsequent filtering before
detection. This alternative description will enable a more
straightforward comparison with the empirical model in
Sec.~\ref{secApp}.

So, instead of the input-output transform Eq.~(\ref{adout}) for
the photon detection operator, we now write
\begin{eqnarray}\label{adoutp}
\hat{a}_{d,{\rm out}}(j)=\vec{\phi}_d^\dag(j)u_f
(u\,\vec{\hat{\text{a}}}+v\,\vec{\hat{\text{a}}}^*),
\end{eqnarray}
where the Bogoliubov matrix $u_f$ accounts for filters (as the
filters are passive, we  have $v_f=0$); this matrix $u_f$ is of
course unitary, even if the filters can present losses. In fact,
losses will be modeled using beamsplitters, where the lost energy
is reflected into auxiliary non-relevant modes. These modes do not
interact with the rest of the experiment (\ie they are unaffected
by Bogoliubov transform $u,v$) and will not reach the APD. But all
the other modes, referred to as relevant modes, should be
considered as detection modes, and we then have:
\begin{equation} \label{equProjPr}
\sum_j\vec\phi_d(j)\vec\phi_d^\dag(j)=\Pi_r,
\end{equation}
where $\Pi_r$ is the projector onto the subspace of relevant
modes. Let $\overline{\Pi}_r$ be the projector on the non-relevant
modes. As the latter are unaffected by the transform $u,v$, we
have $\overline{\Pi}_r
(u\,\vec{\hat{\text{a}}}+v\,\vec{\hat{\text{a}}}^*)=\overline{\Pi}_r\vec{\hat{\text{a}}}$.
Inserting the relation $\Pi_r+\overline{\Pi}_r=\openone$ into
(\ref{adoutp}) then leads to
\begin{eqnarray}\label{adoutp2}
\hat{a}_{d,{\rm out}}(j)=\vec{\phi}_d^\dag(j)u_f\Pi_r
(u\,\vec{\hat{\text{a}}}+v\,\vec{\hat{\text{a}}}^*)+\vec{\phi}_d^\dag(j)u_f\overline{\Pi}_r
\vec{\hat{\text{a}}}.
\end{eqnarray}
Here the last term on the right annihilates vacuum, and commutes
with annihilation operators like $\hat{a}_0$ or $\hat{a}_1$: this
term will add no contribution to the results of the previous
subsection. The only change therefore consists in the substitution
$\vec\phi\rightarrow \Pi_r u_f^\dag\vec\phi$, so that the operator
$\Pi$ in Eq.~(\ref{equProjP}) should be replaced by
\begin{equation} \label{equProjP2}
\Pi'=\Pi_r u_f^\dag\sum_j\vec\phi_d(j)\vec\phi_d^\dag(j) u_f
\Pi_r=F^\dag F,
\end{equation}
where $F=\Pi_ru_f\Pi_r$ and where we have used standard properties
of projection operators ($\Pi=\Pi^\dag$, $\Pi^2=\Pi$). The
operator $F$ represents the action of the filters restricted to
the subspace of relevant modes. If $F$ is a projector, such as
$\Pi_T$ of Eq.~(\ref{equProjPT}) or $\Pi_\Omega$ of
Eq.~(\ref{equProjPOmega}), then the effect of $\Pi'$ is the same
as of $\Pi$ in Eq.~(\ref{equProjP}). This can be easily
understood: it is equivalent to say that the filtered modes are
blocked, or that they are first redirected into auxiliary modes,
and then blocked.

The operator $\Pi'$ of Eq.~(\ref{equProjP2}) is a more general
quantity than $\Pi$ in Eq.~(\ref{equProjP}), however, since $\Pi'$
need not be a projection operator. It can for instance account for
partial absorption of the modes. In that case, the spectral
transmission $T(\omega)$ in Eq.~(\ref{equProjPOmega}) can assume
any value between $0$ and $1$, to account for filtering systems
more complex than a simple spectral slit.

Furthermore, the above expressions can simply be generalized to
situations where several filters are used. For example, if a
spectral slit $\Pi_\Omega$ is followed by a time-domain filter
$\Pi_T$, then the above expressions for $F$ and $\Pi'$ become
$F=\Pi_T\Pi_\Omega$, leading to $\Pi'=\Pi_\Omega \Pi_T\Pi_\Omega$.

\section{Application}\label{secApp}
At this stage, we have a complete description of the final state
starting from the Bogoliubov transform~(\ref{Bogoliubov}). The
results of the previous section are generally valid, since we
started with a multimode model that was left unspecified. Our
purpose now is to establish a concrete link with the
photon-subtraction experiment as described in
Ref.~\cite{Ourjoumtsev2006a}, and to improve its analysis.

\subsection{Photon subtraction experiment}
\begin{figure}[t]
\center
\includegraphics{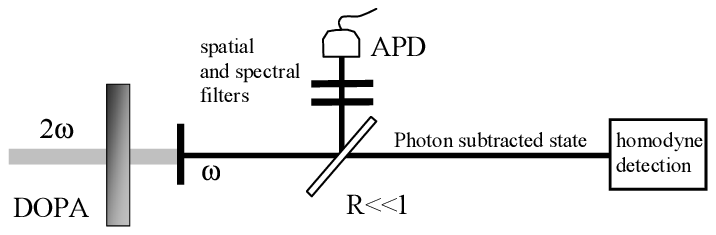}
\caption{Simplified experimental setup: a squeezed vacuum is
generated by a DOPA, where photons of frequency $2\omega$ are
transformed into pairs of photons of frequency $\omega$ (in the
experiment, the central frequency $\omega_0$ corresponds to a
central wavelength of about $850~nm$). The output signal of the
DOPA is sampled by a beam splitter with low reflectivity $R$. If a
photon is detected by the APD, then ideally it has been subtracted
from the squeezed vacuum. } \label{figDOPA}
\end{figure}

 In the experiment by Ourjoumtsev {\em et al.}~\cite{Ourjoumtsev2006a},
pulses of squeezed light are produced. The setup is sketched in
Figure~\ref{figDOPA}.
A squeezed vacuum, produced in a single-pass DOPA (a ${\rm K Nb
O}_3$ crystal) by down-conversion of frequency-doubled femtosecond
laser pulses, is sampled by a beam splitter with low reflectivity
$R = 1-T$. Two mode filters are placed in front of the APD: a
spatial mode filter, that consists in a single-mode fiber, and a
spectral slit of width $\Omega$. If a photon is detected by the
APD, then ideally it has been subtracted from the squeezed vacuum.
This subtraction leads to a $1$-photon squeezed state, which is
very close to a `Schr{\" o}dinger kitten' state. Quantum state
tomography with a balanced homodyne detector~\cite{Leonhardt1997a}
allows the complete reconstruction of this highly non-gaussian
quantum state of light.

The Wigner function~(\ref{Wj}) was derived assuming that the mode
reduction was performed on the mode entering the homodyne
detector. To relate our results to the empirical model discussed
in the next subsection, we here choose to perform the mode
reduction to the signal directly after the DOPA.
We model the DOPA using the scheme presented in Fig.~\ref{modelU},
where the parameters $r$ and $g$ are linked to the Bogoliubov
transform through Eqs.~(\ref{eta}-\ref{equ_gr}). For the
calculation of the modes $\hat{a}_{d,\rm out}(j)$ detected by the
APD, the sampling beamsplitter and the mode filters can be
separately added to this transform, as explained in
Sec.~\ref{subsecDetModes}.
As stated above, here we chose not to include the sampling beam
splitter into the mode reduction. The Wigner function~(\ref{Wj})
then describes the signal just after the DOPA.
We therefore still need to account for this sampling beam splitter
between the DOPA and the homodyne detection, as well as for other
losses.
For example, one usually accounts for imperfections of the
homodyne detection by adding a fictitious beam splitter of
transmission $\eta_{\rm hom}$ just before the homodyne detection,
where $\eta_{\rm hom}$ is the homodyne detection efficiency. Both
those beam splitters can be easily implemented by replacing the
variances according to:
\begin{equation} \label{beamsplit}
V^{(m)}_{x/p}-1=\eta_{\rm hom} T[V_{x/p}-1].
\end{equation}
and by multiplying $A$, $B$ and $D$ by $\eta_{\rm hom} T$. Before
going into detailed calculations, let us first recall the
empirical model that was proposed in Ref.~\cite{Ourjoumtsev2006a}
to account for experimental results.

\subsection{Empirical model}\label{SecEmpirical}

It is useful to  recall the empirical model proposed in
Ref.~\cite{Ourjoumtsev2006a} to explain the experiments, and to
see by what assumptions our multimode model reduces to it. The
DOPA is again modeled as in Fig.~\ref{modelU}, producing the same
squeezed vacuum. However, in the empirical model it is assumed
that the detected photon is either in the homodyne mode with
probability $\xi$, or in an orthogonal mode with probability
$(1-\xi)$. In the latter case, the detection event is not
correlated with the homodyne measurement, and one simply performs
a homodyne measurement on squeezed vacuum.

\begin{figure}[t]
\center
\includegraphics{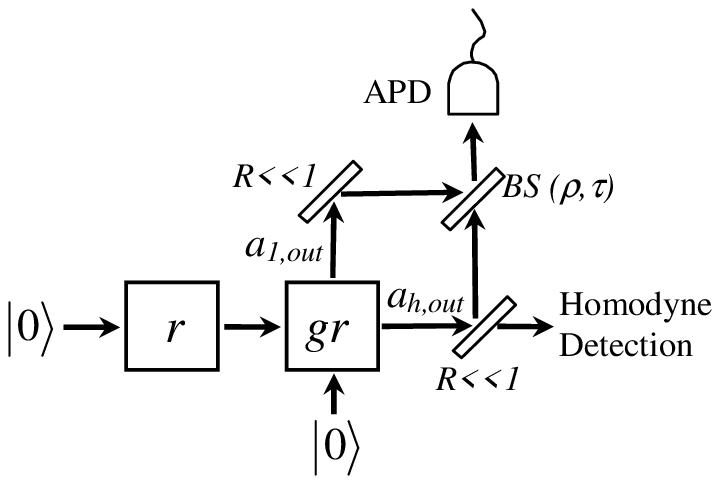}
\caption{When one applies the mode reduction procedure to the
multimode model for photon subtraction as sketched in
Fig.~\ref{genSetup}, the resulting Wigner function is equivalent
to the two-mode model depicted in the figure. The DOPA may in
general be represented by an ideal single mode DOPA and a two-mode
NDOPA as in Fig.~\ref{modelU}, and the photon counting is a
combination of dark counts as well as a coherent mixture of the
two output ports of the NDOPA. Compared to the empirical model
developed in Ref.~\cite{Ourjoumtsev2006a}, the only difference is
the presence of the beam splitter $BS(\rho,\tau)$ which was not
present in the empirical model.} \label{figModelUBeamsplit}
\end{figure}

The output density matrix obtained with the empirical model is
similar to our $\rho_{j}$ in Eq.~(\ref{rho2}). In fact, the two
would be identical
if the detected photon was only due to photons in the mode
$\psi_{h}$ or from $\mathcal{H}_{\perp}$. This is in general not
the case, however, as there will be an admixture from $\hat a_{1,
{\rm out}}$ in the photon detection operator. In fact, in order to
completely account for the multimode nature of this experiment,
the empirical model should be modified in the way depicted on
Fig.~\ref{figModelUBeamsplit}, with the insertion of a
beamsplitter of amplitude reflection and transmission coefficients
$\rho$ and $\tau$ that allows interference between $\hat a_{h,{\rm
out}}$ and $\hat a_{1,{\rm out}}$. A photon detection event in
such a setup can indeed be equivalent to the application of
$\hat{a}_\theta^\dag(j)$ to the initial vacuum [see
Eqs.~(\ref{rho2},\ref{athetaj})], provided
\begin{equation}\label{EquivSetup}
\frac{\rho}{\tau}\cosh(r)=\frac{1}{\tan\theta_j}-\frac{1}{\tan\theta_0},
\end{equation}
with $\tan(\theta_0)=\beta/\alpha$. These angles $\theta_{0}$ and
$\theta_j$ are mixing angles that fix the probability amplitudes
of detection of a photon of mode 0 and of mode 1. Our angle
$\theta_j$ in general depends both on the squeezing properties of
the light source, and on the filtering of the signal before the
photon detector, whereas the empirical $\theta_0$ only depends on
the source. Within a narrow-filter approximation that will be
detailed in next subsection, such a setup can also account for
averaged quantities (\ref{Xbar}) with the use of an average angle
$\bar\theta$ instead of $\theta_j$ (see Eq.~(\ref{theta bar}) in
the following).

This possibility of interference between the homodyne signal and
the $\hat a_{1,{\rm out}}$ signal is the crucial difference
between the multimode and the empirical models: More interference
makes the empirical model worse. The essential assumption of the
empirical model is thus that the photon detection operator does
not have a contribution from $\hat a_{1,{\rm out}}$. Then
$\gamma_j$ and $\delta_j$ could be replaced by $\alpha$ and
$\beta$, respectively, according to
Eqs.~(\ref{modereductionahout}) and (\ref{adperp_Mode_Red}), and
the angle $\theta_j$ in Eq.~(\ref{rho2}) by $\theta_0$ (in which
case, according to Eq.~(\ref{EquivSetup}), the beamsplitter
$BS(\rho,\tau)$ in the equivalent model in
Fig.~\ref{figModelUBeamsplit} can be removed).

The empirical Wigner function can be easily deduced from
Eq.~(\ref{Wj}) with the above replacements, and has the same form
after the replacement of $\bar{A}\cdots\bar{D}$ by $A_{\rm
emp}\cdots D_{\rm emp}$.
Coefficients $A_{\rm emp}$ and $B_{\rm emp}$ are obtained by
multiplying $\bar{A}$, $\bar{B}$ by $\xi/\bar{\xi}$, and replacing
$\bar{\gamma}$ and $\bar{\delta}$ by $\alpha$ and $\beta$,
respectively. This gives
\begin{equation}\label{ABempirical}
A_{\rm emp}=\xi\frac{(V_x-1)^2}{V_x+V_p-2}, \qquad B_{\rm
emp}=\xi\frac{(V_p-1)^2}{V_x+V_p-2}.
\end{equation}
The coefficient $C_{\rm emp}$ is given by $C_{\rm emp}=1-A_{\rm
emp}/V_x-B_{\rm emp}/V_p$, and $D_{\rm emp}$ vanishes.

The empirical model produces intuitive results. However, it
requires justification. If large spectral slits would be used,
then the homodyne mode and many other orthogonal modes would
hardly be affected by the slit. If the detected photon could have
come from many modes orthogonal to $\psi_{h}$, then the modal
purity  $\xi$ would be unacceptably low, and also a large
admixture of $\hat a_{1,{\rm out}}$ would enter the detection
signal. Indeed, some of us found experimentally that the spectral
slit should be as narrow as possible, while still allowing the
detection of a signal, in order to find the highest modal purities
(see also Sec.~\ref{Sec:concretemultimode}). Consequently, narrow
slits have been used in the photon-subtraction
experiment~\cite{Ourjoumtsev2006a}. Although it is obvious that
filtering is necessary, the use of a narrow spectral slit before
the photon detector does not make the empirical model
automatically valid. A quantitative comparison of both models is
therefore needed to test the validity of the empirical model, as
given below.

\subsection{Concrete multimode model}\label{Sec:concretemultimode}
Let us now develop a simple spatio-temporal multimode model for
which the Bogoliubov transformation can be written explicitly. We
assume that light propagation inside the DOPA is described by
modes of the form
\begin{equation} \label{SVEA0}
A(\vec{r},t)\exp(i\omega_0 t-i\vec{k}\cdot\vec{r}),
\end{equation}
where the plane wave is exactly phase-matched, and where the
amplitude $A$ satisfies the slowly-varying envelope approximation
(SVEA). This approximation does not hold for all the light that
exits the nonlinear crystal, but the homodyne mode is supposed to
be phase-matched, and we will assume that the filters before the
APD block the modes that are not phase-matched. We will
furthermore neglect diffraction effects within the DOPA. In the
basis $(x,y,t)$, where $x$ and $y$ are spatial variables running
on the DOPA's transverse plane, the $u$ and $v$ of the Bogoliubov
transform~(\ref{Bogoliubov}) then become diagonal (see appendix
\ref{AppendixBogoliubov}), and we have
\begin{equation} \label{Bogoliubov2}
\hat{\text{a}}_{\rm
out}(x,y,t)=u(x,y,t)\hat{\text{a}}(x,y,t)+v(x,y,t)\hat{\text{a}}^\dag(x,y,t),
\end{equation}
in terms of operators that we assume to have commutation relations
\begin{equation}
[\hat a(x,y,t),\hat a^{\dag}(x',y',t')]=
\delta(x-x')\delta(y-y')\delta(t-t').
\end{equation}
Coefficients in the transformation~(\ref{Bogoliubov2}) have the
form
\begin{subequations}\label{UVxyt}
\begin{eqnarray}
u(x,y,t)  & = & \cosh[\,q l \, E_P(x,y,t)\,], \label{Uxyt}\\
v(x,y,t) & = & \sinh[\,q l \, E_P(x,y,t)\,]. \label{Vxyt}
\end{eqnarray}
\end{subequations}
Here $E_P(x,y,t)$ is the pump-beam amplitude, which we assume to
be real-valued. The parameter $q$ takes into account the
nonlinearity of the crystal and $l$ is its length. One can allow
for Group Velocity Mismatch (GVM) in the crystal by convoluting
$E_P$ by a rectangular unit gate of duration $\tau_g$, the time
separation induced by the GVM after passing the crystal (see
appendix \ref{AppendixBogoliubov}). More precisely, this
convolution should be made twice, as we also have GVM for the
Second Harmonic Generation (SHG) of the pump beam. The  $u$ and
$v$ are real-valued functions if the pump beam $E_P$ is so, which
is a valid assumption if there is no frequency chirp. The homodyne
mode $\psi_h(x,y,t)\equiv e_h(x,y,t)$ will also be taken
real-valued in the following.

The homodyne signal is then given by $\hat a_{h,{\rm out}}=\int
e_{h} \hat{\text a}_{\rm out}$, where integration over $x,y,t$ is
implied. Mode reduction now starts with the identification
\begin{equation}\label{etaforpulses}
\eta\, \hat a_{0} = \int \mbox{d}x\mbox{d}y\mbox{d}t\,
e_{h}(x,y,t) u(x,y,t) \hat{\text a}(x,y,t),
\end{equation}
which is Eq.~(\ref{a0}) specified for our spatiotemporal model.
The mode-reduction parameters are now given by spatio-temporal
integrals, for example
\begin{subequations}\label{alphaeta2}
\begin{eqnarray}
\eta^2 & = & \int \mbox{d}x\mbox{d}y\mbox{d}t\, e_h^2(x,y,t)\,u^2(x,y,t) \label{eta2}\\
\eta\alpha & = & \int \mbox{d}x\mbox{d}y\mbox{d}t\,
e_h^2(x,y,t)\,u(x,y,t)\,v(x,y,t). \label{alpha2}
\end{eqnarray}
\end{subequations}
Further parameters can be found analogously.

\smallsection{Averaging over photon detection events} We have seen
in Sec.~\ref{subsecDetModes} that all averaged quantities can be
obtained through the determination of the operator $\Pi$ defined
in (\ref{equProjP2}) when the filters are separately added to the
Bogoliubov transform. In the considered experiment two filters are
used: a rectangular spectral slit, that can be described using
(\ref{equProjPOmega}); and a monomode fiber that selects a single
spatial mode $\phi_s(x,y)$, which we suppose to be real-valued,
and therefore corresponds to the projector
$\phi_s(x,y)\phi_s(x',y')$. Note that in this experiment detection
times are unknown at the scale of pulses duration, so that there
is no time-domain filtering.  (In the analysis one should average
over all possible photon detection times.) We therefore have to
use
\begin{equation} \label{equPmodel}
\Pi(x,y,\omega;x',y',\omega')=\phi_s(x,y)\phi_s(x',y')T(\omega)\delta(\omega-\omega'),
\end{equation}
where $T(\omega)=\eta_c R$ if $\omega$ enters into the spectral
slit, $0$ otherwise. Here $R$ is the sampling beamsplitter
reflectivity, and $\eta_c$ accounts for all other losses in the
conditioning arm (APD efficiency, optics losses ...). As can be
seen from Eq.~(\ref{SVEA0}), the field amplitudes are defined
around a central frequency $\omega_0$ (or $2\omega_0$ for the DOPA
pump beam, see appendix~\ref{AppendixBogoliubov}), so that the
 frequency $\omega=0$ for the amplitude Fourier transform
corresponds in fact to this central frequency; in that way, a
rectangular spectral slit well centered around this central
frequency can be defined as $\omega\in[-\Omega/2,\Omega/2]$. The
operator $\Pi$ defined in Eq.~(\ref{equPmodel}) should be applied
to Fourier-transformed mode functions:
\begin{equation} \label{fourier}
\tilde{\phi}(\omega)=\int\mbox{d}t\, \phi(t)e^{-\mathrm{i} \omega
t}.
\end{equation}
Using Eqs.~(\ref{Pj}),~(\ref{equPmodel}) and~(\ref{fourier}), the
average total photon detection probability per pulse then becomes
\begin{equation} \label{Ptotcalc}
P_{\rm
tot}=\frac{\eta_cR\Omega}{4\pi^2}\int\mbox{d}x\mbox{d}y\mbox{d}t\,
v^2(x,y,t)\phi_s^2(x,y),
\end{equation}
where the average is taken over the detection modes
$\phi_{d,j}(x,y,t)$. The expression~(\ref{Ptotcalc}) increases
linearly with the filter width $\Omega$, but is valid only if
$\Omega$ is small enough to warrant the SVEA. Furthermore, when
conditioning upon a click in the detector in
Sec.~\ref{SecConditioning}, we assumed that $P_{\rm tot} \ll 1$,
an assumption that can now be tested with the explicit
formula~(\ref{Ptotcalc}).

In general, the parameters  $\gamma$ and $\delta$ appear in
$\overline{W}$ as product sums like
$\overline{\gamma\gamma^*}=\sum_j|\gamma_j|^2$, or
$\overline{\gamma\delta^*}=\sum_j\gamma_j\delta_j^*$. In our
concrete model, the evaluation of detection-averaged coefficients
in the Wigner function involves integrals of the type:
\begin{equation} \label{integral}
\frac{\eta_cR}{4\pi^2}\int\mbox{d}x\mbox{d}y\,\mbox{d}x'\mbox{d}y'\int_{-\Omega/2}^{\Omega/2}\mbox{d}\omega\,
\tilde{f}(x,y,\omega)\tilde{h}^*(x',y',\omega).
\end{equation}
For example, the average $\overline{\gamma\gamma^*}$ is found by
substituting both $\tilde{f}$ and $\tilde{h}$ in~(\ref{integral})
by the time-domain Fourier transform of the function $e_h\, u\,
v\,\phi_s (x,y,t)$. [Here and in the following, we abbreviate
products like $f(x,y,t)h(x,y)$ by $f h (x,y,t)$.] Since $\tilde f$
and $\tilde h$  are Fourier transforms of real-valued functions,
the corresponding integrals~(\ref{integral}) are real-valued as
well. Hence, all mode-reduction parameters and coefficients in the
Wigner functions are also real-valued. In particular, the
coefficient $\bar{D}$ in $\bar{W}$ vanishes, see Eq.~(\ref{Dj}).
There is no difficulty to numerically evaluate
integrals~(\ref{integral}) and we will do that below, but let us
first focus on an additional approximation that can considerably
simplify these results, without becoming inaccurate.

\smallsection{Narrow-filter approximation}
We previously discussed the experimental observation that the
spectral slit should be as narrow as possible. Another
simplification is possible in that case, that simply consists in
neglecting in the integrals~(\ref{integral}) the frequency
dependence of the mode profiles within the narrow width $\Omega$,
{\em i.e.} $\tilde{f}(x,y,\omega) \simeq \tilde{f}(x,y,0)$ for
$|\omega| < \Omega/2$. Let us recall that this value $\omega=0$
corresponds to the central frequency $\omega_0$ of the pulses,
where the real-valued amplitudes present a maximum. This has two
consequences: first, the presence of this maximum justifies a
zero-order Taylor approximation, provided the spectral width
$\Omega$ is much smaller than the spectral width of the pulses
(typically $2\pi/\tau$, where $\tau$ is the pulse duration);
second, if all functions involved in~(\ref{integral}) present a
maximum at $\omega=0$, then the narrow-filter approximation
generates an upper-bound for these integrals, and therefore for
quantities like $\overline{|\gamma^2|}$, $\overline{|\delta^2|}$
or the modal purity $\overline{\xi}$ (see~\ref{xij},\ref{Xbar}).
This approximation will be applied and tested in
Sec.~\ref{SecNumerical}, dedicated to the numerical results. This
approximation brings the following simplification in the
integrals~(\ref{integral}):
\begin{eqnarray} \label{simple}
&=&\frac{\eta_cR\Omega}{4\pi^2}
\int\mbox{d}x\mbox{d}y\,\tilde{f}(x,y,0)\int\mbox{d}x'\mbox{d}y'\,\tilde{h}^*(x',y',0)\\
&=&\frac{\eta_cR\Omega}{4\pi^2} \int\mbox{d}x\mbox{d}y\mbox{d}t\,
f(x,y,t)\int\mbox{d}x'\mbox{d}y'\mbox{d}t'\, h^*(x',y',t').
\nonumber
\end{eqnarray}
Evidently, we end up with separate integrals over $f$ and $h$, and
using the definitions~(\ref{gammadelta}) we obtain the averages
\begin{subequations}\label{gammabdeltab}
\begin{eqnarray}
\bar{\gamma} & =&
\frac{\sqrt{\eta_cR\Omega}}{2\pi}\frac{1}{\eta}\int\mbox{d}x\mbox{d}y\mbox{d}t\,
e_h\,u\,v\,\phi_s (x,y,t)
\label{gammab}\\
\bar{\delta} & = & -\frac{\alpha\bar{\gamma}}{\beta} +
\frac{1}{\beta}\frac{\sqrt{\eta_cR\Omega}}{2\pi}\int
\mbox{d}x\mbox{d}y\mbox{d}t\, e_h\,v^2\phi_s (x,y,t).
\label{deltab}
\end{eqnarray}
\end{subequations}
In the narrow-filter approximation, averages of products are
simply given by products of averages,
$\overline{\gamma\gamma^*}=|\bar{\gamma}|^2$, and
$\overline{\gamma\delta^*}=\bar{\gamma}\bar{\delta}^*$, etc.
Essentially in the $\Omega \rightarrow 0$ limit the filter removes
any temporal information about the time the photon was emitted
from the DOPA. The photodetection then corresponds to a single
mode with $\omega = 0$, regardless of the average over detection
times.

We therefore find for the photon-subtracted squeezed state an
average Wigner function of the form~(\ref{constantsinW}), with
coefficients
\begin{subequations}\label{AbarBbar}
\begin{eqnarray}
\bar{A} & = & P_{\rm tot}^{-1}|\bar{\gamma}(\eta+\alpha)+\bar{\delta}\beta|^2\label{Abar}\\
\bar{B} & = & P_{\rm
tot}^{-1}|\bar{\gamma}(\eta-\alpha)-\bar{\delta}\beta|^2,\label{Bbar}
\end{eqnarray}
\end{subequations}
and $\bar{C}=1-\bar{A}/V_x-\bar{B}/V_p$ and $\bar{D}=0$. Using the
same substitution in (\ref{xij},\ref{athetaj}) we can also
introduce the averaged modal purity
\begin{equation}\label{ksi bar}
\bar{\xi}=\frac{|\bar{\gamma}|^2+|\bar{\delta}|^2}{P_{\rm tot}}
\end{equation}
and the average angle $\bar{\theta}$ defined by:
\begin{equation}\label{theta bar}
\tan\bar{\theta}=\bar{\delta}/\bar{\gamma}.
\end{equation}

\smallsection{Constant profiles}
Before dealing with a more realistic case, it is interesting to
focus on the case of constant profiles. Let us assume a constant
value for $e_h$, $E_P$, and $\phi_s$ within a space-time support
of volume $\Xi$. The normalization of the homodyne mode implies
$e_h^2=1/\Xi$, and equation~(\ref{UVxyt},\ref{alphaeta2}) leads to
$\eta=u$, $\alpha=v$, and $\beta=0$. As $\beta$ vanishes, the mode
$\hat{a}_1$ is no more defined, and
Eqs.~(\ref{delta},\ref{deltab}) cannot be used anymore. In fact
the mode-reduction procedure now leads to an effective {\em
single}-mode model rather than a two-mode model. The homodyne mode
is now in the single-mode space $\mathcal{H}_1$ spanned by
$\hat{a}_0$,$\hat{a}_0^\dag$. One can simply put $\delta_j=0$
in~(\ref{adperp_Mode_Red}), and hence $\bar\delta=0$. This leads
to the average angle $\bar{\theta}=\theta_0=0$. The modal purity
becomes $\bar{\xi}=1$, as it should for a single-mode model. Most
importantly, we find  $\bar{C}=-1$, which according to
Eq.~(\ref{constantsinW})  corresponds to the most negative value
for the Wigner function at the origin, $W(0,0) = -1/\pi$.

So, with constant profiles and a narrow filter slit, we recover
from our multimode model the single-mode description for photon
subtraction experiments. Since this limit leads to the most
negative Wigner function, it represents the ideal limit for
producing states for QIP applications, at least according to our
simple multimode model. This shows that the multimode nature
essentially appears through the mode distortions due to the
non-constant space and time profiles of the pulses. The central
role of gain-induced distortions is particularly clear with regard
to the multimode nature of the squeezed vacuum produced by the
DOPA: assuming a constant pump field, that is assuming no
gain-induced distortions, is in fact enough to obtain $\beta=0$.
Let us now return to a more realistic model, taking into account
these profiles.

\smallsection{Gaussian profiles} As a more realistic
simplification, we assume gaussian profiles for the various
fields. For instance, we write the homodyne field $e_h$ as
\begin{equation}
e_{h}(x,y,t)=e_{h,0}\exp\left(-\frac{x^2+y^2}{w^2}\right)\exp\left(-2\frac{t^2}{\tau^2}\right),
\end{equation}
where $w$ is the beam waist, $\tau$ the duration of the gaussian
pulse, and $e_{h,0}$ a normalization constant. The pump beam is
usually obtained by SHG, in a crystal pumped by a beam identical
to the homodyne beam. In the lowest order of the SHG process, the
profiles of $E_P$ and $e_h^2$ have the same shapes, so we can
assume another gaussian profile:
\begin{eqnarray}
E_P(x,y,t) & = & E_0 e_P(x,y,t)\label{EPxyt}\\
 & = & E_0\exp\left(-\frac{x^2+y^2}{w_P^2}\right)\exp\left(-2\frac{t^2}{\tau_P^2}\right),\nonumber
\end{eqnarray}
where one expects the pump pulse duration to be $\tau_P=\tau/\sqrt
2$. However, if GVM is taken into account, this gaussian
profile~(\ref{EPxyt}) must be convoluted by rectangular gates. In
practice, such convolutions lead to  beam profiles that are still
very close to gaussians.

The final gaussian profile to be introduced here is the spatial
mode $\phi_s(x,y)$ of the filter in front of the APD. It is the
${\rm LP}_{01}$ mode  of a monomode fiber that is well
approximated by a normalized gaussian of waist $w_f$.

\subsection{Numerical results}\label{SecNumerical}
Here our goal is twofold: first, to compare our multimode analysis
with the empirical model that was used before to analyze photon
subtraction experiments. Second, by exploring our multi-parameter
multimode model, we look for parameter regimes that are best
suited for producing states with the most negative Wigner
functions.

\smallsection{Expansions in pump field} For our numerical work it
is convenient to write all fields as Taylor expansions in the pump
field $E_P$. From Eq.~(\ref{UVxyt}) it follows directly that
\begin{subequations}\label{UUUVVV}
\begin{eqnarray}
u^2(x,y,t) & = &  \frac{1}{2} + \frac{1}{2}\cosh[2 q l E_{P}] = \sum_m b_m e_P^m, \\
u\,v(x,y,t) & = & \frac{1}{2}\sinh[2 q l E_{P}] = \sum_m c_m e_P^m,  \\
v^2(x,y,t) & = &  - \frac{1}{2} + \frac{1}{2}\cosh[2 q l E_{P}] =
\sum_m d_m e_P^m,
\end{eqnarray}
\end{subequations}
with $E_{P} = E_{0} e_P(x,y,t)$ as in Eq.~(\ref{EPxyt}).  This
defines the constant coefficients $b_{m},c_{m}$, and $d_{m}$. For
$q l E_0 < 1$ these expansions converge quite quickly. We then
only have to insert relations (\ref{UUUVVV}) into the various
integrals for an efficient numerical evaluation. For instance, we
can rewrite (\ref{eta2}) as:
\begin{equation}\label{eta2Numeric}
\eta^2=\sum_m b_mP_m
\end{equation}
with
\begin{equation}\label{Pm}
P_m=\int \mbox{d}x\mbox{d}y\mbox{d}t\,
e_h^2e_P^m=\frac{2\sqrt{2}w_P^2\tau_P}{(mw^2+2w_P^2)\sqrt{m\tau^2+2\tau_P^2}}.
\end{equation}
In the same way we have
\begin{subequations}\label{Numeric}
\begin{eqnarray}
\eta\alpha=\sum_m c_mP_m, \label{etaAlphaNumeric}\\
P_{\rm tot}=\eta_cR\Omega\sum_m d_mQ_m, \label{PtotNumeric}\\
\bar{\gamma}=\frac{\sqrt{\eta_cR\Omega}}{\eta}\sum_m c_mR_m,
\label{gammaNumeric}\\
\bar{\delta}=-\frac{\alpha\bar{\gamma}}{\beta}+\frac{\sqrt{\eta_cR
\Omega}}{\beta}\sum_m d_mR_m, \label{deltaNumeric}
\end{eqnarray}
\end{subequations}
with
\begin{subequations}\label{Numeric2}
\begin{eqnarray}
Q_m &=&\frac{1}{4\pi^2}\int\mbox{d}x\mbox{d}y\mbox{d}t\, \phi_s^2e_P^m=\frac{\pi^{-3/2}\tau_Pw_P^2}{2\sqrt{2m}(mw_{\rm f}^2+2w_P^2)}\label{Qm}\\
R_m &=&\frac{1}{2\pi}\int \mbox{d}x\mbox{d}y\mbox{d}t\, \phi_se_P^me_h\\
     &=& \frac{\pi^{-3/4}}{\sqrt{\tau}\sqrt{\tau^{-2}+m\tau_P^{-2}}}\frac{1}{ww_{\rm f}(w^{-2}+mw_P^{-2}+w_{\rm f}^{-2})}.\nonumber
     \label{Rm}
\end{eqnarray}
\end{subequations}
After fixing parameters, these expansions in the pump field can be
readily used for numerical evaluations.

\smallsection{Fixing basic parameters} First we fix some
parameters of our multimode model in order to present numerical
results and to see how much our analysis differs from the one in
Ref.~\cite{Ourjoumtsev2006a}, where filtering before the photon
detection was not modeled explicitly. We take $w=1.2w_P$ and a
transmission $T=90\%$ of the sampling beam splitter. Moreover, we
fix  $w_{\rm f}=w/1.5$, which is compatible with the coupling
efficiency into the filtering monomode fiber (approximately
$80\%$, see Ref.~\cite{Ourjoumtsev2006a}).

Regarding efficiency of homodyne detection,  the mode $\hat{a}_h$
considered in Sec.~\ref{secModered} was defined as the mode that
perfectly matches the local oscillator of the homodyne detection,
in other words the matching efficiency equals unity by definition
in our model.
The transmission of the optics and the photodetection efficiency
together lead to an overall efficiency of homodyne detection
$\eta_{\rm hom}$. We put $\eta_{\rm hom}=0.93$, in agreement with
Ref.~\cite{Ourjoumtsev2006a}.

As stated above, if GVM is taken into account, the almost gaussian
profile of the pump pulse is convoluted twice by a rectangular
gate with time window $\tau_{g}$. A ${\rm K Nb O}_3$ crystal of
length $l= 100\,{\rm \mu m}$ has $\tau_g=120\,{\rm fs}$. For an
initial duration of the homodyne pulse $\tau\approx 150\,{\rm
fs}$, the convolutions indeed lead to a nearly gaussian beam
profile with $\tau_P\approx \tau$. We assume the identity $\tau_P=
\tau$ in the following.

\begin{figure}[t]
\begin{center}
\includegraphics{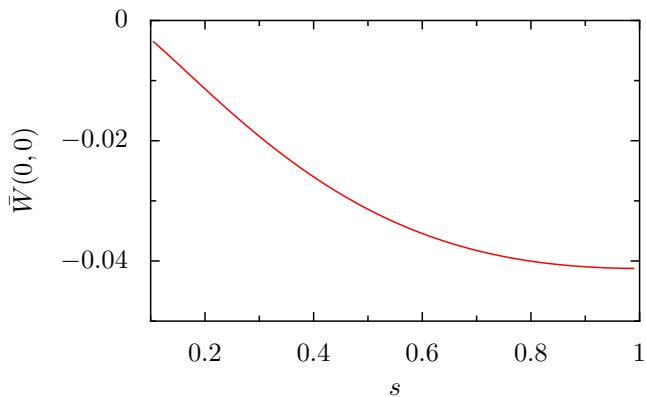}
\end{center}
\caption{(color online). Minimal value of the Wigner function,
$\bar{W}(0,0)$, as a function of squeezing parameter
$s=\exp(-2r)$, which is varied by changing the quantity $qlE_{0}$.
The narrow-filter approximation was made for the spectral slit.
Fixed parameters: pulse parameters $w = 1.2 w_{P}$, $\tau_{P}=\tau
= 150\,{\rm fs}$, transmission of sampling beam splitter
$\mathrm{T}= 90\%$, efficiency of homodyne detection $\eta_{\rm
hom} = 0.93$. } \label{W0fs}
\end{figure}

\smallsection{Negative Wigner functions} As stated in the
Introduction, the global minimum of a Wigner function is the
standard figure of merit for the nonclassicality and
`non-gaussianity' of the corresponding state. After subtraction of
a single photon, the Wigner function $\bar{W}(x,p)$ is always most
negative in the origin (since $\bar{D}= 0$). Figure~\ref{W0fs}
shows how $\bar{W}(0,0)$ depends on the squeezing factor
$s=\exp(-2r)$. The most negative values are obtained in the
low-squeezing limit $s \rightarrow 1$. This can be understood as
there is less gain-induced distortions in that case.
\begin{figure}[t]
\begin{center}
\includegraphics{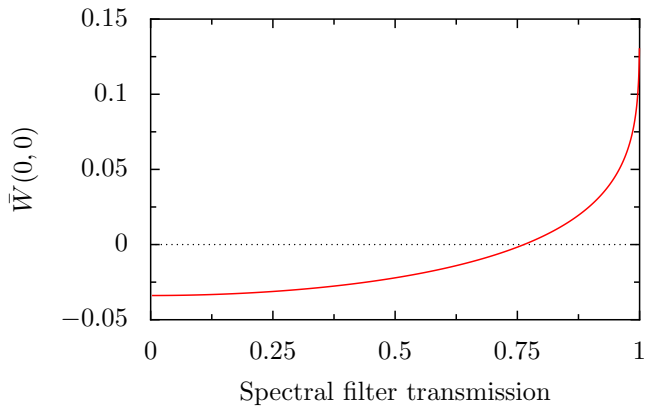}
\end{center}
\caption{(color online). Minimal value of the Wigner function,
$\bar{W}(0,0)$, as a function of the spectral slit transmission
for the homodyne mode. Calculated for $s=0.56$, using a complete
evaluation of integrals ~(\ref{integral}), i.e. without the
narrow-filter approximation. The narrow-filter approximation is
well satisfied for low transmissions.} \label{W0ModelCompl}
\end{figure}

In Ref.~\cite{Ourjoumtsev2006a}, the best experimental results
(highest modal purities) were obtained  for $s=0.56$. For this
value of $s$, which can be selected by choosing the right value
for the quantity $qlE_{0}$, we obtain $g=0.50$ and
$\bar{W}(0,0)=-0.034$; the latter value is close to what was
observed in~\cite{Ourjoumtsev2006a}, without correction for the
detection efficiency. At this stage, it can be interesting to
compare this result, obtained using the narrow-filter
approximation, with a more accurate calculation based on a
complete evaluation of integrals ~(\ref{integral}).
Figure~\ref{W0ModelCompl} presents the numerical results obtained
for $\bar{W}(0,0)$ at $s=0.56$ as a function of the spectral slit
transmission for the homodyne mode. (This transmission can be
increased by making the spectral slit width $\Omega$ larger.)  A
minimal value is clearly reached for low transmissions, justifying
{\it a posteriori} the use of narrow spectral slits in the
experiment of Ref.~\cite{Ourjoumtsev2006a}. Since for low
transmissions, $\bar W(0,0)$ does not differ much from its minimal
value, the narrow-filter approximation that we made in
Sec.~\ref{Sec:concretemultimode} gives accurate results.

\begin{figure}[t]
\begin{center}
\includegraphics{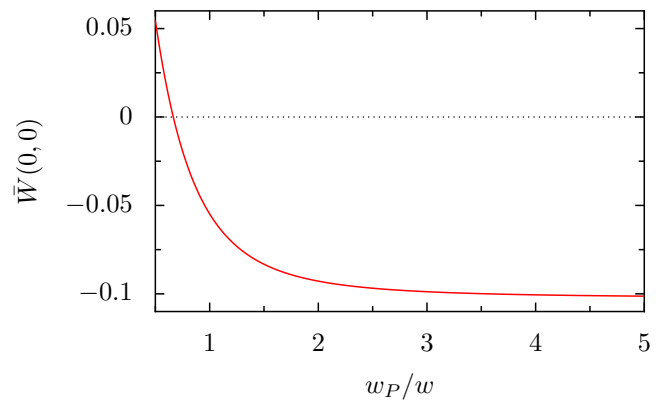}
\end{center}
\caption{(color online). Minimal value of the Wigner function, \ie
$\bar{W}(0,0)$, as a function of $w_P/w$, where $w_P$ is the waist
of the pump field and $w$ is the waist of the homodyne field.
$qlE_0$ is fixed such that squeezing parameter $s=0.56$. Other
parameter values as in Fig.~\ref{W0fs}.} \label{W0fw}
\end{figure}
Figure~\ref{W0fw} predicts the behavior of $\bar{W}(0,0)$ when
varying the size of the pump beam. Experimental values for the
widths were related by $\omega = 1.2
\omega_{P}$~\cite{Ourjoumtsev2006a}. Fig.~\ref{W0fw} clearly shows
that one can await a high increase of the negativity from a larger
pump beam. This result was intuitive, as there is less
gain-induced distortions in that case, but is here quantified.
This can motivate the use of amplified pulses~\cite{Dantan}, in
order to have a spatially broader pump beam ({\em i.e.} with
larger $w_{P}$), but with the same intensity.

\begin{figure}[t]
\begin{center}
\includegraphics{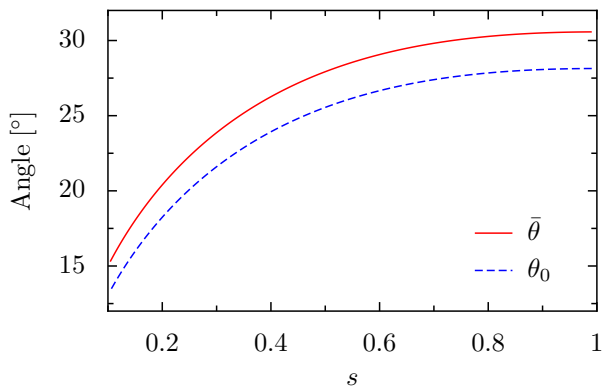}
\end{center}
\caption{(color online). Mixing angles $\bar\theta$ (solid line)
and $\theta_0$ (dashed line) of modes 0 and 1, as a function of
squeezing parameter $s$, which is varied by changing $qlE_{0}$.
Other parameter values as in Fig.~\ref{W0fs}. } \label{theta}
\end{figure}

\smallsection{Comparison with empirical model} Measured negative
Wigner functions were interpreted in~\cite{Ourjoumtsev2006a} using
the empirical model as introduced in Sec.~\ref{SecEmpirical},
where the photon is subtracted in the `good' mode $\hat{a}_{h,{\rm
out}}$ with probability $\xi$, and where the state is left in the
squeezed vacuum with probability $(1-\xi)$. As explained before,
the main difference between the empirical and our models is that
the conditioned state in the former corresponds to the
single-photon initial state $\hat{a}^\dag_{\theta_0}$ with
$\tan\theta_0=\beta/\alpha$, while it corresponds to
$\hat{a}^\dag_{\bar{\theta}}$ in the latter, with
$\tan\bar{\theta}=\bar{\delta}/\bar{\gamma}$. These angles
$\theta_{0}$ and $\bar \theta$ are mixing angles that fix the
probability amplitudes of detection of a photon of mode 0 and of
mode 1. Figure~\ref{theta} shows $\theta_0$ and $\bar\theta$ as a
function of the squeezing parameter $s$.
Clearly, $\theta_0$ and $\bar\theta$ do not differ too much, and
by less than $10\%$ for $s=0.56$.

Another important difference between our model and the empirical
model is that modal purities $\xi_{j}$ in our model are fixed by
the relation~(\ref{xij}), whereas the parameter $\xi$ in the
empirical model is a free parameter. This freedom can be used to
fit the data, \ie to have $A_{\rm emp}=\bar{A}$ or $B_{\rm
emp}=\bar{B}$. It is however not {\em a priori} possible to fit
both parameters $\bar{A}$ and $\bar{B}$ (\ref{Abar},\ref{Bbar})
from (\ref{ABempirical}) using only the fitting parameter $\xi$.
In neither model should the variances $V_x$ and $V_p$ be
considered as free fitting parameters of the photon-subtraction
experiment, at least their values should agree with the values for
$V_{x,p}$ obtained by homodyne measurements of the squeezed
vacuum.

\begin{figure}[t]
\begin{center}
\includegraphics{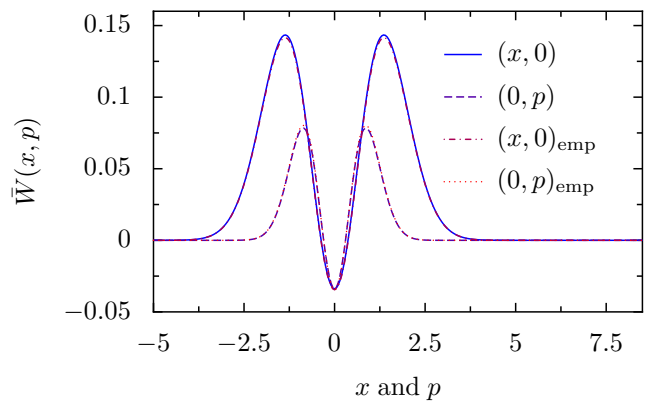}
\end{center}
\caption{(color online). Sections along $(x,0)$ and $(0,p)$ of the
Wigner function $\bar{W}(x,p)$ (solid line), and corresponding
best fits using the empirical model (dashed line) with $\xi_{\rm
opt}=0.87$. Parameter values as in Fig.~\ref{W0fs}. Results of our
model and the empirical fits almost overlap. } \label{figWigner}
\end{figure}

In our model the mixing angles $\theta_{j}$ and their average
$\bar \theta$ take into account the filtering of the signal that
is used for conditioning. In the empirical model, the
corresponding angle $\theta_{0}$ is independent of the filtering.
Thus it is to be expected that this inaccuracy of the empirical
model will lead to optimally fitted modal purities $\xi_{\rm opt}$
in the empirical model that are systematically lower than the
average modal purity $\bar \xi$ in our model. This is indeed what
we find for the curves in Figure~\ref{figWigner}: the best fit in
the present example is obtained for $\xi_{\rm opt}=0.87$, a value
that is indeed smaller but still close to  $\overline{\xi}\approx
0.91$. The high quality of this fit (with an error less than
$1.2\%$) is directly linked to the fact that in the present case
$\theta_0\approx\overline{\theta}$.

There is a possibility to improve this result if $g$ is considered
as a fitting parameter as well. We obtained an error of less than
$0.4\%$ between the Wigner functions for  $\xi_{\rm opt}=0.89$ and
$g_{\rm opt}=0.53$, \ie for a value of $g$ that differs by $6\%$
from the value given by the multimode model. In other words, if
$g$ has a great influence on $A_{\rm emp}$ and $B_{\rm emp}$ in
(\ref{ABempirical}), it has a very low impact on $V_x$, $V_p$; the
change from $g=0.50$
 to $g_{\rm opt}=0.53$ modifies the values of $V_x$, $V_p$
by only a few $10^{-3}$, and for this reason it is very difficult
to accurately measure $g$ from squeezed vacuum \cite{Wenger}.
These considerations explain why the empirical model can fit
experimental data so successfully; even when $\theta_0$ is not
equal to $\bar{\theta}$, the parameter $g$ gives a supplementary
freedom for fitting.

\section{Discussion and conclusions}\label{secDiscConc}
We have introduced a straightforward and physically intuitive
procedure that we call `mode reduction' to simplify the multimode
description of squeezed light to the bare essentials. For
photon-subtraction experiments, this means that the homodyne
signal is reduced to an effective two-mode description and the
detector signal requires one extra orthogonal effective mode. We
derived the Wigner function of the homodyne signal conditional
upon the detection of a single photon, and we also showed how to
average over possible measurement outcomes.


The general mode-reduction formalism was then applied to a
detailed model describing photon subtraction of gaussian
spatiotemporal pulses of squeezed light. This model features many
experimental parameters such as beam waists and duration of the
pulses that can be independently measured. Indeed, our model does
not have free fitting parameters. This allows one to study in
detail what are the crucial experimental parameters to produce
optimally negative Wigner functions with pulses of squeezed light.

We compared our new model to the empirical model that was used
before to analyze photon-subtraction experiments in
\cite{Ourjoumtsev2006a}. In fact, the formulae for the output
Wigner functions look similar. One crucial difference is that the
empirical model does have a free parameter, namely the quantity
called the modal purity. In our model modal purities also occur,
be it with a slightly different meaning, but they are fixed
quantities. A good agreement between our model and experiments
therefore gives more understanding than an accurate fit with the
empirical model.

We found that in the range of parameters of the measurements
in~\cite{Ourjoumtsev2006a}, both our model and the empirical model
are accurate. We reasoned that modal purities in our model would
be systematically higher, and in our numerical example we found
this to be the case. The accuracy of the empirical model strongly
depends on the availability of the free parameter. It was
nevertheless a surprise in the theoretical analysis that the
mixing angles $\theta_{0}$ and $\bar \theta$, describing the
relative probability of measuring a photon in either one of two
effective modes, differ at most $20 \%$ in a whole range of
squeezing parameters.

Our mode-reduction procedure is closely related to the analysis of
photon-subtraction experiments of
Refs.~\cite{Molmer,Nielsen2007a,Nielsen2007b}. One could express
our mode-reduction parameters in terms of elements of the
covariance matrix of
Refs.~\cite{Molmer,Nielsen2007a,Nielsen2007b}. Our output Wigner
function in Eq.~(\ref{Wj}) then reduces to the one in
Ref.~\cite{Nielsen2007b}, but only in the special case that all
our mode reduction parameters are  real-valued so that $D_{j}$ in
Eq.~(\ref{Dj}) vanishes. This we assumed for simplicity in
Sec.~\ref{Sec:concretemultimode}. Our mode-reduction procedure  is
carried out in the Heisenberg picture. We think that our approach
has some advantages.  In our approach it becomes quite intuitive
in what sense it goes beyond the empirical model of
Ref.~\cite{Ourjoumtsev2006a}. In our concrete multimode analysis,
we include effects not considered in Ref.~\cite{Nielsen2007b},
such as  the transverse beam profile, for which we found that
wider pump beams lead to more negative Wigner functions.

In conclusion, we presented a very concise model that can account
for the multimode nature of projective photon-counting
measurements. It gives an intuitive picture of photon-subtraction
experiments, close to the empirical model previously published.
This multimode model therefore gives consistent results, in
agreement with previously published experiments where pulses of
light with negative Wigner functions were produced conditionally.
Our model can be used to predict the changes in the output upon
variation of experimentally relevant parameters, and to optimize
the setup design.

\begin{acknowledgments}
We thank K. M{\o}lmer for useful discussions. This work has been
supported by the Danish Research Council through QUANTOP, by
COMPAS, and by the Niels Bohr International Academy.
\end{acknowledgments}

\appendix

\section{Wigner function} \label{AppendixWigner}

In order to derive Eq.~(\ref{Wj}) from Eq.~(\ref{Winitial}), one
has first to invert Eqs.~(\ref{modereductionahout},\ref{a1out}),
leading to
\begin{eqnarray}
\hat{a}_0 & = &\eta(\hat{a}_{h,{\rm
out}}-\frac{\beta\hat{a}^\dag_{1,{\rm
out}}}{\sqrt{1+\beta^2}})-\alpha(\hat{a}^\dag_{h,{\rm
out}}-\frac{\beta\hat{a}_{1,{\rm out}}}{\sqrt{1+\beta^2}})\label{invA0}\\
\hat{a}_1 & = &\sqrt{1+\beta^2}\hat{a}_{1,{\rm
out}}-\beta\hat{a}^\dag_{h,{\rm out}}. \label{invA1}
\end{eqnarray}
One should then replace in Eq.(\ref{Winitial}) $x_{0,1},p_{0,1}$
by $x_{h,{\rm out}}, p_{h,{\rm out}}, x_{1,{\rm out}}$, and
$p_{1,{\rm out}}$, and integrate over $x_{1,{\rm out}}$,
$p_{1,{\rm out}}$. It is however convenient to make a change of
variables so that the integral is over $x_1$, $p_1$ instead of
$x_{1,{\rm out}}$, $p_{1,{\rm out}}$. In this case the only
transforms needed for this calculation is
\begin{subequations}\label{rel_x0p0}
\begin{eqnarray}
x_0 &= &\frac{\eta-\alpha}{1+\beta^2}(x_{h,{\rm out}}-\beta x_1)\label{rel_x0}\\
p_0 & = & \frac{\eta+\alpha}{1+\beta^2}(p_{h,{\rm out}}+\beta p_1),\\
\nonumber\label{rel_p0}
\end{eqnarray}
\end{subequations}
as well as the transformation of the integral
\begin{equation}\label{x1p1integral}
\int \mbox{d}x_{1,{\rm out}} \mbox{d}p_{1,{\rm out}} =
\frac{1}{1+\beta^2} \int \mbox{d}x_1 \mbox{d}p_1.
\end{equation}
One should then note that the Wigner function~(\ref{Winitial}) is
the product of a polynomial in $x$, $p$, and of a gaussian term
$\exp{(-R^2)}$, with
\begin{equation}
R^2=x_0^2+p_0^2+x_1^2+p_1^2=x_{\theta}^2+p_{\theta}^2+x_{\theta+\pi/2}^2+p_{\theta+\pi/2}^2.
\end{equation}
With Eq.~(\ref{rel_x0p0}), the exponent can be rewritten as
\begin{eqnarray}
R^{2}& = &\frac{V_x}{(\eta+\alpha)^2}\left(x_1-\frac{\beta
x_{h,{\rm
out}}}{V_x}\right)^2+\frac{x_{h,{\rm out}}^2}{V_x}\nonumber\\
&+& \frac{V_p}{(\eta-\alpha)^2}\left(p_1-\frac{\beta p_{h,{\rm
out}}}{V_p}\right)^2+\frac{p_{h,{\rm out}}^2}{V_p}.\nonumber
\end{eqnarray}
The integral~(\ref{x1p1integral}) with Eq.~(\ref{Winitial}) as its
integrand can then be found by replacing in the integrand the
squares $x_1^2$ and $p_1^2$ by
\begin{subequations}
\begin{eqnarray}
x_1^2\rightarrow\frac{\beta^2 x_{h,{\rm
out}}^2}{V_x^2}+\frac{(\eta+\alpha)^2}{2V_x}\\
p_1^2\rightarrow\frac{\beta^2 p_{h,{\rm
out}}^2}{V_p^2}+\frac{(\eta-\alpha)^2}{2V_p},
\end{eqnarray}
\end{subequations}
and by replacing the first-order terms according to
\begin{subequations}
\begin{eqnarray}
x_1\rightarrow\frac{\beta x_{h,{\rm
out}}}{V_x}\\
p_1\rightarrow\frac{\beta p_{h,{\rm out}}}{V_p}.
\end{eqnarray}
\end{subequations}

\section{Slowly Varying Envelope Approximation}
\label{AppendixBogoliubov} The goal of this appendix is the
derivation of the local Bogoliubov
transformation~(\ref{Bogoliubov2}). We assume that inside the DOPA
the pump pulse with an angular frequency $2\omega_0$  travels at a
speed $v_{g,2\omega_0}$, with negligible absorption. This field
can therefore be written as
\begin{equation}
\mathrm{i}\,E_P(x,y,t-\frac{z}{v_{g,2\omega_0}}-\delta
t)\exp(2i\omega_0 t-i\vec{k}_{2\omega_0}\cdot\vec{r}),
\end{equation}
where $\delta t$ is an arbitrary time delay and where
`$\mathrm{i}$' is a purely conventional phase factor. Let us write
the probe beam as
\begin{equation}
E(\vec{r},t)=A(\vec{r},t)\exp(i\omega_0
t-i\vec{k_{\omega_0}}\cdot\vec{r}),
\end{equation}
where the phase-matching condition
$\vec{k}_{2\omega_0}=2\vec{k}_{\omega_0}$ is assumed to be
satisfied. By using the SVEA in Maxwell's equations, neglecting
diffraction terms and considering the first-order dispersion, we
obtain
\begin{equation}
\frac{\partial A}{\partial
z}+\frac{1}{v_{g,\omega_0}}\frac{\partial A}{\partial t}=q
E_P(x,y,t-\frac{z}{v_{g,2\omega_0}}-\delta t)A^*,
\end{equation}
where $A=A(\vec{r},t)$. The substitution of $t-z/v_{g,\omega_0}$
by $t$ then leads to
\begin{equation}\label{SVAAeq}
\frac{\partial A}{\partial z}(\vec{r},t)=q E_P(x,y,t-Dz-\delta
t)A^*(\vec{r},t),
\end{equation}
where $D=v^{-1}_{g,2\omega_0}-v^{-1}_{g,\omega_0}$ is the GVM.
With the assumption that $E_P$ is real-valued, the solution to
Eq.~(\ref{SVAAeq}) becomes
\begin{equation}\label{Aoutsolved}
A_{\rm out}=\cosh[\,q l F_P \,]\,A_{\rm in}+\sinh[\,q l F_P
\,]\,A^*_{\rm in},
\end{equation}
where the $(x,y,t)$-dependence was suppressed. The effective pump
field $F_{P}$ is given by
\begin{equation}\label{Fpintegral}
F_P=\frac{1}{l}\int_0^l dz E_P=\frac{1}{\tau_g}\int_{\delta
t}^{\tau_g+\delta t} d\tau E_P(x,y,t-\tau),
\end{equation}
with  $\tau_g=Dl$ the time separation induced by the GVM after
crossing the crystal. Eq.~(\ref{Fpintegral}) shows that the
effective pump field $F_{P}$ is a convolution of the pump field
$E_P$ with  a rectangular unit gate of duration $\tau_g$, which
for $\delta t=-\tau_g/2$ is centered around the origin. As the
quantized version of Eq.~(\ref{Aoutsolved}), we then find
Eq.~(\ref{Bogoliubov2}) of the main text. The pump field $E_{P}$
of the main text is to be understood as the effective pump field
$F_{P}$ derived here. Evidently, $F_{P} \rightarrow E_{P}$ in the
limit $\tau_{g}\rightarrow 0$ (no GVM).




\begin{references}


\bibitem{Eisert2002a} J.~Eisert, S.~Scheel, and M.B.~Plenio, Phys. Rev. Lett. {\bf 89}, 137903 (2002).
%
\bibitem{Giedke} G.~Giedke and J.I.~Cirac, Phys. Rev. A {\bf 66}, 032316 (2002).
%
\bibitem{Browne} D.E.~Browne, J.~Eisert, S.~Scheel, and M.B.~Plenio, Phys. Rev. A {\bf 67}, 062320 (2003).
%
\bibitem{Kim2005a} M.S.~Kim, E.~Park, P.L.~Knight, and H.~Jeong, Phys. Rev. A {\bf 71}, 043805 (2005).
%
\bibitem{Biswas2007a} A.~Biswas and G.S.~Agarwal, Phys. Rev. A {\bf 75}, 032104 (2007).
%
\bibitem{Dakna1997a} M.~Dakna, T.~Anhut, T.~Opatrn{\' y}, L.~Kn{\" o}ll, and \\ D.-G.~Welsch, Phys. Rev. A {\bf 55}, 3184 (1997).
%
\bibitem{Opatrny2000a} T.~Opatrn{\' y}, G.~Kurizki, and D.-G.~Welsch, Phys. Rev. A {\bf 61}, 032302 (2000).
%
\bibitem{Lvovsky} A.I.~Lvovsky, H.~Hansen, T.~Aichele, O.~Benson, J.~Mlynek, and  S.~Schiller,
Phys. Rev. Lett. {\bf 87}, 050402 (2001).
%
\bibitem{WengerCond} J.~Wenger, R.~Tualle-Brouri and P.~Grangier, Phys. Rev. Lett. {\bf 92}, 153601
(2004).
%
\bibitem{Ourjoumtsev2006a} A.~Ourjoumtsev, R.~Tualle-Brouri, J.~Laurat, and Ph.~Grangier, Science {\bf 312}, 83 (2006), and Supporting
    Online Material.
%
\bibitem{Ourjoumtsev2006b} A.~Ourjoumtsev, R.~Tualle-Brouri, and Ph.~Grangier, Phys. Rev. Lett. {\bf 96}, 213601 (2006).
%
\bibitem{Wakui} K.~Wakui, H.~Takahashi, A.~Furusawa and M.~Sasaki, e-print quant-ph/0609153v1.
%
\bibitem{Neergaard} J.S.~Neergaard-Nielsen, B. Nielsen, C.~Hettich, K.~M{\o}lmer, and E.S.~Polzik, Phys. Rev. Lett. {\bf 97},
    083604 (2006).
%
\bibitem{Ourjoumtsev2006c} A.~Ourjoumtsev, H.~Jeong, R.~Tualle-Brouri, P.~Grangier, Nature {\bf 448}, 784 (2007).
%
\bibitem{Parigi2007a} V.~Parigi, A.~Zavatta, M.~Kim, and M.~Bellini, Science {\bf 317}, 1890 (2007).
%
\bibitem{Kim2008a} M.S.~Kim, J. Phys. B: At. Mol. Opt. Phys. {\bf 41}, 133001 (2008).
%
\bibitem{Avenhaus} M. Avenhaus, H. B. Coldenstrodt-Ronge, K. Laiho, W. Mauerer, I. A.
Walmsley and C. Silberhorn, Phys. Rev. Lett. {\bf 101}, 053601
(2008).
%
\bibitem{Suzuki} M.~Sasaki and S.~Suzuki, Phys. Rev. A {\bf 73}, 043807 (2006).
%
\bibitem{Molmer} K.~M{\o}lmer, e-print quant-ph/0602202v1.
%
\bibitem{Nielsen2007a} A.E.B.~Nielsen and K.~M{\o}lmer, Phys. Rev. A {\bf 75}, 023806 (2007).
%
\bibitem{Nielsen2007b} A.E.B.~Nielsen and K.~M{\o}lmer, Phys. Rev. A {\bf 76}, 033832 (2007).
%
\bibitem{prl} F.~Grosshans and Ph.~Grangier,  Phys. Rev. Lett. {\bf 88}, 057902 (2002).
%
\bibitem{Aichele} T. Aichele, A.I. Lvovsky and S. Schiller, Eur. Phys. J. D {\bf 18}, 237
(2002).
%
\bibitem{LaPorta} A. La Porta and R. E. Slusher, Phys. Rev. A {\bf 44}, 2013 (1991).
%
\bibitem{Gardiner2000a} C.W.~Gardiner and P.~Zoller, {\em Quantum Noise} (Springer, Berlin, 2000).
%
\bibitem{Wenger} J.~Wenger, J.~Fiur\'a\v{s}ek, R.~Tualle-Brouri,
N.J.~Cerf, and P.~Grangier, Phys. Rev. A \textbf{70}, 053812
(2004).
%
\bibitem{Leonhardt1997a}U.~Leonhardt, {\em Measuring the Quantum state of Light} (Cambridge University Press, 1997).
%


\bibitem{Dantan} A.~Dantan, J.~Laurat, A.~Ourjoumtsev, R.~Tualle-Brouri, and P.~Grangier, Optics Express {\bf 15}, 8864 (2007).
\end{references}
\end{document}